\documentstyle[12pt,epsfig]{article}
\def\lsim{\raise0.3ex\hbox{$<$\kern-0.75em\raise-1.1ex\hbox{$\sim$}}}
\def\gsim{\raise0.3ex\hbox{$>$\kern-0.75em\raise-1.1ex\hbox{$\sim$}}}

\def\pom{{I\!\!P}}

\def\beq{\begin{equation}}
\def\eeq{\end{equation}}
\def\bea{\begin{eqnarray}}
\def\eea{\end{eqnarray}}
\def\bq{\begin{quote}}
\def\eq{\end{quote}}

\def\gappeq{\mathrel{\rlap {\raise.5ex\hbox{$>$}}
{\lower.5ex\hbox{$\sim$}}}}

\def\lappeq{\mathrel{\rlap{\raise.5ex\hbox{$<$}}
{\lower.5ex\hbox{$\sim$}}}}

\def\Toprel#1\over#2{\mathrel{\mathop{#2}\limits^{#1}}}

\def\pom{{I\!\!P}}

\begin{document}
\pagestyle{empty}
\begin{center}
{\bf DOUBLE MESON PRODUCTION IN ULTRAPERIPHERAL HEAVY ION COLLISIONS }
\\
\vspace*{1cm}
 V.P. Gon\c{c}alves $^{1}$, M.V.T. Machado  $^{1,\,2}$\\
\vspace{0.3cm}
{$^{1}$ Instituto de F\'{\i}sica e Matem\'atica,  Universidade
Federal de Pelotas\\
Caixa Postal 354, CEP 96010-090, Pelotas, RS, Brazil\\
$^{2}$ \rm High Energy Physics Phenomenology Group, GFPAE,  IF-UFRGS \\
Caixa Postal 15051, CEP 91501-970, Porto Alegre, RS, Brazil}\\
\vspace*{1cm}
{\bf ABSTRACT}
\end{center}
 
\vspace*{1mm}
\noindent
The double meson production in ultraperipheral heavy ions collisions is addressed, focusing on the particular case of $\rho\, J/\Psi$ from two-photon reactions. The cross section at photon level is obtained using distinct parameterizations for the gluon distribution on the light meson. The resulting  estimates for the nuclear case are presented and discussed. As a by product, we estimate the double $\rho$ production cross section  using the Pomeron-exchange factorization relations. 
\vspace*{1cm}
\noindent
\rule[.1in]{16.5cm}{.002in}

\vspace{-2cm}
\setcounter{page}{1}
\pagestyle{plain}

\vspace{2cm}

Relativistic heavy-ion collisions are a potentially prolific
source of $\gamma \gamma$ collisions at high energy colliders  \cite{reviewbaur,bert}.
 The
advantage of using heavy ions is that the cross sections varies as
$Z^4 \alpha^4$ rather just as $\alpha^4$ as in the $e^+e^-$ collisions. Moreover, the maximum
$\gamma \gamma$ collision energy,  $W_{\gamma \gamma}$, is $2\gamma
/R_A$,  about 6 GeV at RHIC and 200 GeV at LHC, where $R_A$ is the
nuclear radius and $\gamma$ is the center-of-mass system (cms)  Lorentz
factor of each ion. In particular, the LHC will have a significant
energy and luminosity reaching beyond LEP2, and could be a bridge to
$\gamma \gamma$ collisions at a future $e^+ e^-$ linear collider.
For the large values of energy at LHC 
 the hard QCD Pomeron is
presumably the dominant mechanism of production. Consequently, the estimate of the cross sections should to consider the QCD 
 dynamics effects. 
Recently, we have proposed in Refs. \cite{vicmag,vicmag2}
  to investigate
QCD Pomeron effects  in
photon-photon scattering at ultraperipheral heavy ion collisions.
In particular, in Ref. \cite{vicmag} we have analyzed the 
diffractive double $J/\Psi$ production in $\gamma \gamma$ collisions,
with the photons coming from the Weizs\"acker - Williams spectrum
of the nuclei. For that process  our results have indicated that
future experimental analyzes  can be useful to discriminate the
QCD dynamics at high energies. Here, we extend the analysis of double meson
production for the $\rho J/\Psi$ case  using a perturbative approach   and provide reliable estimates for the cross sections concerning that
reaction. Since the Pomeron-exchange factorization theorem \cite{gribov,block} allow us to obtain the $\rho \, \rho$ cross section in terms of the $\rho J/\Psi$ and $J/\Psi J/\Psi$ cross sections, we use our results for these cross sections  to estimate the double $\rho$ production in ultraperipheral heavy ion collisions. 
 Moreover, since  the double meson  production can also  occur in photo-nucleus reactions  if the
multiple interactions are considered, we compare our predictions for the two-photon process with that presented in Ref.  \cite{klein} for the photonuclear case.

Let us start considering  the $\rho J/\Psi$ production  process at the photon level, that is $\gamma \gamma \rightarrow \rho\,J/\Psi$, with almost real photons. In the calculations presented here one follows the pioneering  work in Ref. \cite{motykaziaja}. The goal of the present work is also twofold: investigate a discrimination  among models for the gluon distribution on the light meson in an enhanced nuclear cross section and providing reliable estimates for the ultraperipheral nuclear cross section. The differential cross section is estimated in  a similar way as the elastic $J/\Psi$ photoproduction off the proton \cite{Ryskin}. Therefore, it reads as \cite{motykaziaja},
\begin{eqnarray}
\frac{d \sigma \,(\gamma \gamma \rightarrow \rho\, J/\Psi)}{dt}\,(W_{\gamma \gamma}^2,t) = {\cal C}\,\,\alpha_{em}\,g_{\rho}^2 \,\frac{16\,\pi^3 \,[\alpha_s(M_{J/\Psi}^2/4)]^2 \,\,\Gamma_{ee}^{J/\Psi}}{3\,\alpha_{em}\,M_{J/\Psi}^5}\, [\,xG^{\rho}(x,M_{J/\Psi}^2/4)\,]^2\,\, \exp\left( B_{\rho\,J/\Psi}\,t\right)\,\,,
\label{sigmat}
\end{eqnarray}
where ${\cal C}$ denotes factors of corrections discussed later on. The two-photon cms energy is denoted by $W_{\gamma \gamma}$, where $x=M_{J/\Psi}^2/W_{\gamma \gamma}^2$ and $M_{J/\Psi}$ is the heavy meson mass. In the small-$t$ approximation, the slope is estimated to be $B_{\rho\,J/\Psi}=5.5 \pm 1.0$ GeV$^{-2}$ (See further  discussion  and Ref. \cite{motykaziaja}). The light meson-photon coupling is denoted by $g_{\rho}^2=0.454$ and the heavy meson decay width into a  lepton pair is $\Gamma_{ee}$.

\begin{figure}[t]
\centerline{\psfig{file=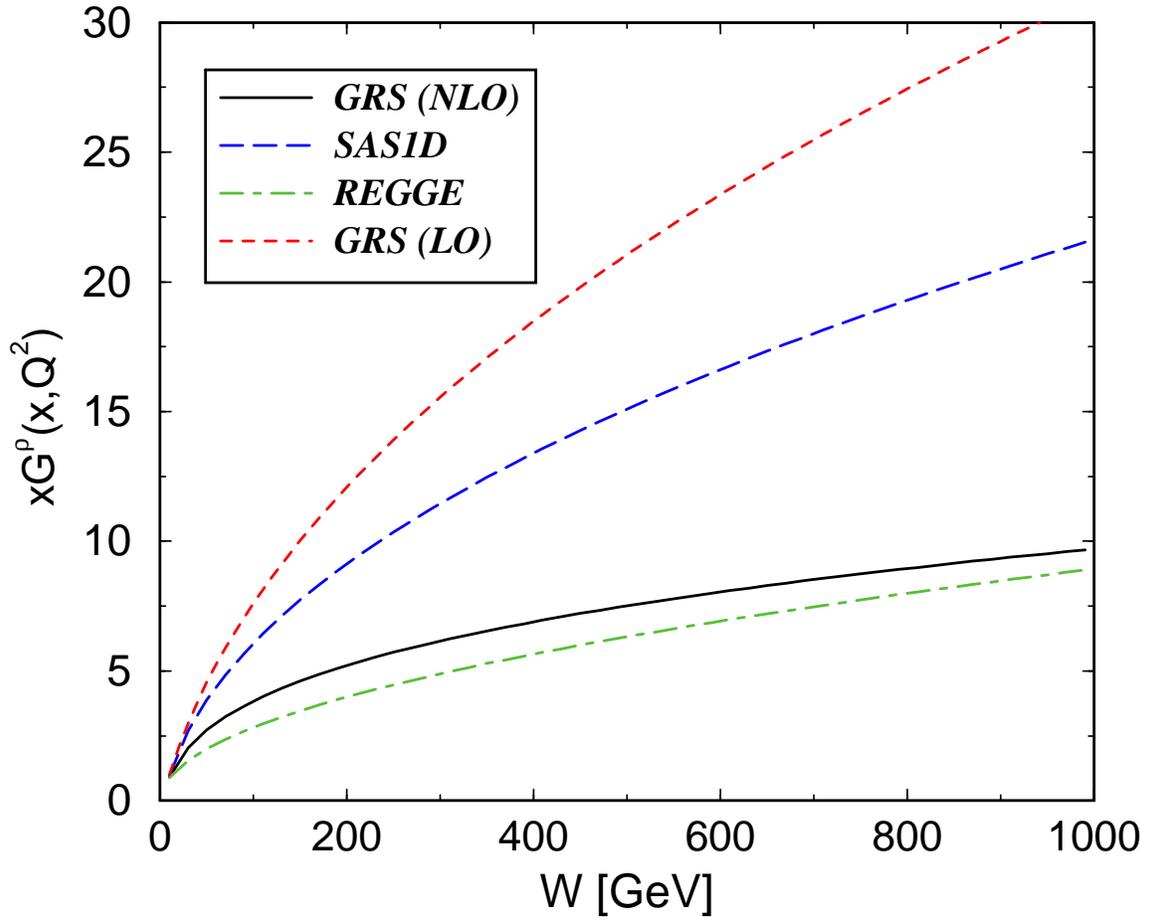,width=150mm}}
 \caption{The gluon distribution on the light  meson for different parameterizations. The solid and dashed lines are NLO and LO GRS, respectively. The long dashed line denotes the SaS1D parameterization and the dot-dashed one the Regge motivated ansatz. }
\label{fig1}
\end{figure}

The process above was proposed as a probe of the gluon distribution on the meson $xG^{rho}$ and, by consequence, a constrain for the photon structure. The enhancement in the sensitivity  by taking the square of those distributions in the total cross section  could discriminate them in measurements at the future photon colliders. In Fig. \ref{fig1}, we present the distributions for the parameterizations considered here as a function of two-photon cms energy. They are the following:  the LO (dashed line) and NLO (solid line) GRS \cite{GRS} parameterization; the SaS1D parameterization \cite{SAS} (long-dashed line) and the phenomenological Regge motivated ansatz (dot-dashed line) proposed at Ref. \cite{motykaziaja}. The latter is given by,
\begin{eqnarray}
x\,G^{rho} (x,M^2_{J/\Psi}/4) = x_0\,G^{rho} (x_0, M^2_{J/\Psi}/4) \,\,\left(\frac{x_0}{x} \right)^{\omega_{\pom}-1}\,,
\label{regge}
\end{eqnarray}
with $x_0=0.1$  and the Pomeron intercept is considered as $\omega_{\pom}=1.25$, in agreement with the effective power in the HERA data. In further analysis for the nuclear case we allow for a higher intercept, in order to simulate a BFKL-like behavior, motivated by the studies in the double $J/\Psi$ production \cite{vicmag}. The normalization, $x_0\,G^{rho} (x_0, M^2_{J/\Psi}/4 )$, is given by the NLO-GRS parameterization.
The differences  among the parameterizations are sizeable as a consequence of distinct effective exponent $\lambda$. The GRS LO one presents the steeper behavior, followed by  SaS1D ($\lambda=0.3038$). The Regge motivated and GRS NLO parameterizations are more close since the intercept for the  Regge ansatz is similar to the effective power of GRS NLO ($\lambda \simeq 0.228$). These behaviors will produce distinct results for the total cross section at photon level and in the nuclear case, allowing discrimination in future colliders.

Let us now calculate the total cross section at photon level considering the models for the gluon distribution on the light meson referred above. In order to do so, we address the correction factor ${\cal C}$. It accounts for several improvements to the process  as partons skewness (off-diagonal) \cite{MartinRyskin}, QCD NLO corrections to the $\gamma J/\Psi$ impact factor \cite{RRML} and real part of the amplitude. Following \cite{motykaziaja}, one has ${\cal C}=C_{\mathrm{off}}\,C_{\mathrm{NLO}}\,C_{\mathrm{rp}}$ where the correction factors  are taken as,
\begin{eqnarray}
C_{\mathrm{off}} & \simeq  & 1.2 \,, \hspace{2cm} C_{\mathrm{NLO}}  \simeq    1+ \frac{\alpha_s\,(M^2_{J/\Psi}/4)}{2}\,,\\
C_{\mathrm{rp}} & \simeq & 1 + \frac{\pi^2\,\lambda}{4}\,, \hspace{1cm} \lambda = \frac{\partial \log \,[\,xG(x,Q^2)\,]}{\partial \log (1/x)}\,. 
\end{eqnarray}

\begin{figure}[t]
\centerline{\psfig{file=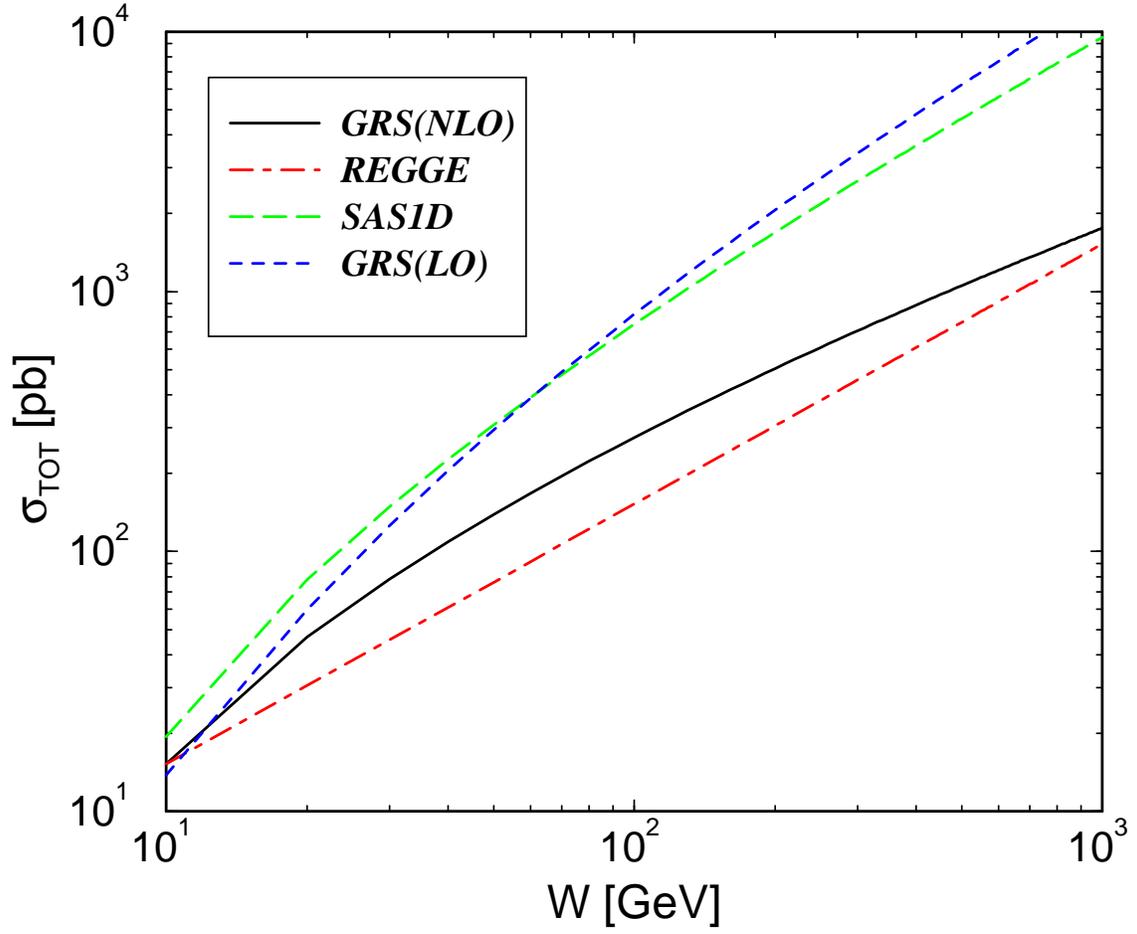,width=150mm}}
 \caption{The total cross section for the process $\gamma \gamma \rightarrow \rho\, J/\Psi$ as a function of two-photon the cms energy. The distinct curves correspond to the different parameterizations.}
 \label{fig2}
\end{figure}

In Fig. \ref{fig2}  we show our predictions for the  total cross section $\sigma_{tot}(\gamma \gamma \rightarrow \rho\,J/\Psi)$ as a function of two-photon cms energy, extrapolated up to 1 TeV. The behavior on energy of the total cross section is strongly dependent of the choice of the gluon content of the light meson. In particular, a strong enhancement of the cross section is predicted if we assume the  LO GRS parameterization, $C_{\mathrm{NLO}}=1$ and disregard    the real part of amplitude. The other predictions are similar to the   results obtained in  \cite{motykaziaja}. It is worth mentioning that this process has been also calculated in Ref. \cite{donnachie}, considering a two-Pomeron model supplemented by the dipole-dipole picture. There, the result is similar to the Regge ansatz and lower than the other parameterizations considered here.

\begin{figure}[t]
\centerline{\psfig{file=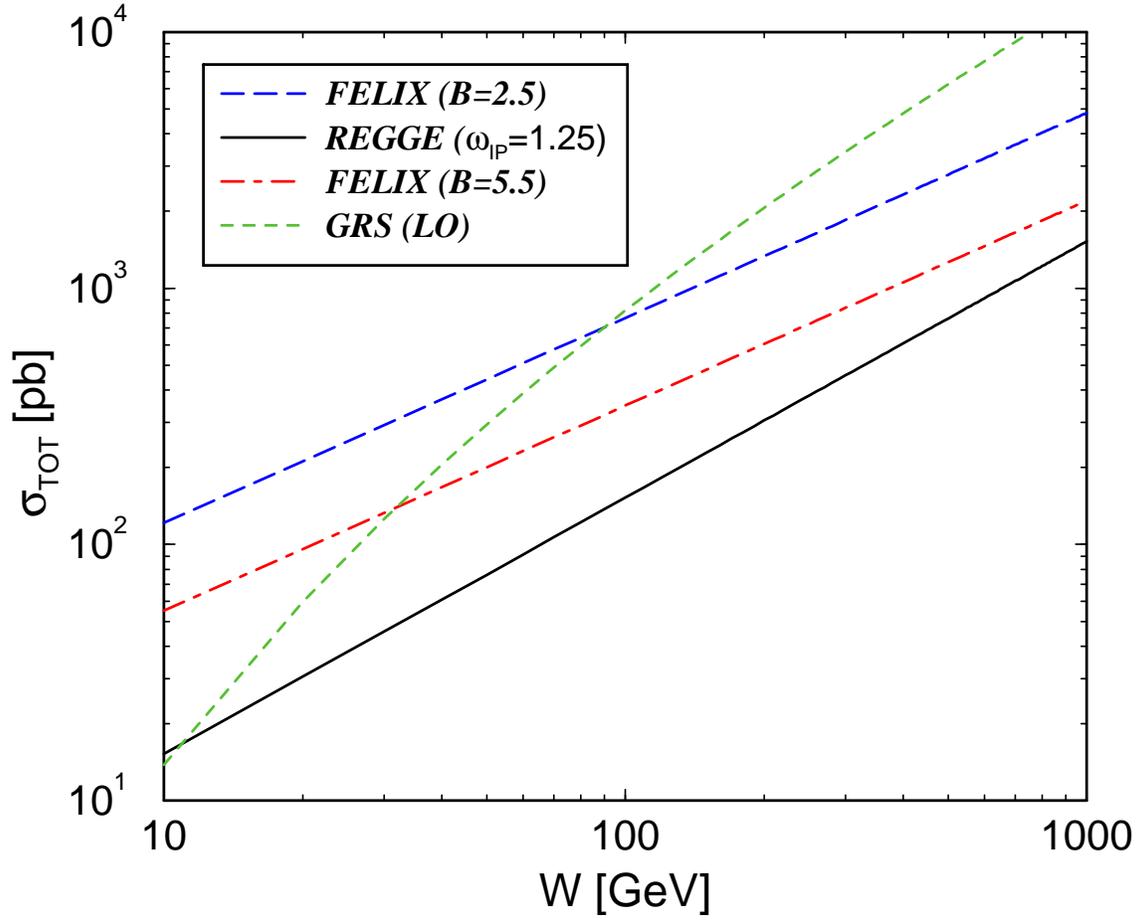,width=150mm}}
 \caption{The same as Fig. \ref{fig2} considering the FELIX anzatz for the cross section and different values of the slope. For comparison our predictions using the Regge motivated input is also presented. }
 \label{fig3}
\end{figure}

Recently, the FELIX Collaboration has proposed the construction of a full acceptance detector for the LHC, with a primary proposal  providing comprehensive observation of a very broad range of strong-interaction processes. In particular, studies of two photon physics in $AA$ collisions has been discussed in its proposal \cite{felix}. There, the differential cross-section $d\sigma (\gamma \gamma \rightarrow V_1 V_2) / dt$, with $V_i = \rho, \, J/\Psi\, ...$,  was parameterized in the form:
\begin{eqnarray}
\frac{d\sigma}{dt} (\gamma \gamma \rightarrow V_1 V_2) = A_{V_1 V_2} \, \left(\frac{W}{W_0}\right)^{C_{V_1 V_2}} \, \exp \left(t\, B_{V_1 V_2} + 4 \,t \alpha^{\prime}_{V_1  V_2} \ln \frac{W}{W_0}\right) \,\,, \label{felixeq}
\end{eqnarray}
where $A_{V_1 V_2} = 1.1$ nb/GeV$^{-2}$, $B_{V_1 V_2} = 2.5$  GeV$^{-2}$, $C_{V_1 V_2} = 0.8$ and $\alpha^{\prime}_{V_1 V_2} = 0$ for the $\rho J/\Psi$ case. 
In Fig. \ref{fig3} we compare the FELIX prediction (long-dashed line)  with our results obtained considering the Regge motivated and GRS(LO) gluon parameterizations as input. Moreover, since in this paper we are considering $B_{\rho J/\Psi} = 5.5$ GeV$^{-2}$, we also present the FELIX prediction for the cross section if we assume this value for the slope.  We see that there is a reasonable agreement between the distinct predictions for the cross section at the photon level.  
Our conclusion at this level is that  the forthcoming photon colliders could  experimentally to check our predictions,   which will  shed light on the vector meson and photon structure.

One comment is in order here. Our value for the slope is corroborated by the recent results for $\rho$ and $J/\Psi$ production at HERA and the proof of the factorization hypothesis for the slope parameters \cite{block}, which predicts that 
\begin{eqnarray}
B_{\gamma \gamma \rightarrow \rho J/\Psi} (W)  = \frac{B_{\gamma p \rightarrow \rho p} (W) \times B_{\gamma p \rightarrow  J/\Psi p} (W)}
{B_{pp}(W)} \,\,,
\label{facbs}
\end{eqnarray}
where $B_{\gamma p \rightarrow \rho p}$,  $B_{\gamma \gamma \rightarrow J/\Psi p}$ and  $B_{pp}$ are the slope of the diffractive peak for the reactions $\gamma  p \rightarrow \rho p$, $\gamma p \rightarrow J/\Psi p$ and $pp \rightarrow pp$, respectively. Data on $\gamma p \rightarrow V p$ in the energy range  $50 \, \mathrm{GeV} < W < 100 \mathrm{GeV}$  gives $B_{\gamma  p \rightarrow \rho p} \approx 11 \, \mathrm{GeV}^{-2}$ and  $B_{\gamma  p \rightarrow J/\Psi p} \approx 5 \,$ GeV$^{-2}$. From $pp$  reactions one obtains  $B_{  p p} = 10 - 12 \,$ GeV$^{-2}$. Consequently, assuming the validity of the factorization we obtain that  $B_{\gamma  \gamma \rightarrow \rho J/\Psi} \approx 5.5 \,$ GeV$^{-2}$ is predicted, which is in agreement with the ansatz proposed in \cite{motykaziaja}.

We are now ready to compute the double meson production in ultraperipheral heavy ions collisions. For two-photon reactions, the cross section for the process $AA \rightarrow AA \,\rho\, J/\Psi$  will be written as,
\begin{eqnarray}
\sigma_{AA \rightarrow AA \rho\, J/\Psi}(s) = \int d\tau \, \frac{d
{\cal{L}}_{\gamma \gamma}}{d\tau}\,(\tau) \, \hat \sigma_{\gamma \gamma
\rightarrow \rho\, J/\Psi}\,(\hat s), \label{sigfoton}
\end{eqnarray}
where $\tau = {\hat s}/s$, $\hat s = W_{\gamma \gamma}^2$ is the square of the cms energy of the two photons and $s$
of the ion-ion system,  $\frac{d\, {\cal{L}}_{\gamma \gamma}\,(\tau)}{d\tau}$  
is the  two-photon  luminosity
 and $\hat \sigma_{\gamma \gamma \rightarrow
\rho \, J/\Psi}(\hat s)$ is the cross section of the $\gamma
\gamma$ interaction (For details related to the numerical
expressions for the effective photon luminosity,  see Ref. \cite{vicmag}). Our approach excludes possible final state interactions of the produced particles with the colliding particles, allowing reliable calculations of
ultraperipheral heavy ion collisions.

\begin{figure}[t]
\centerline{\psfig{file=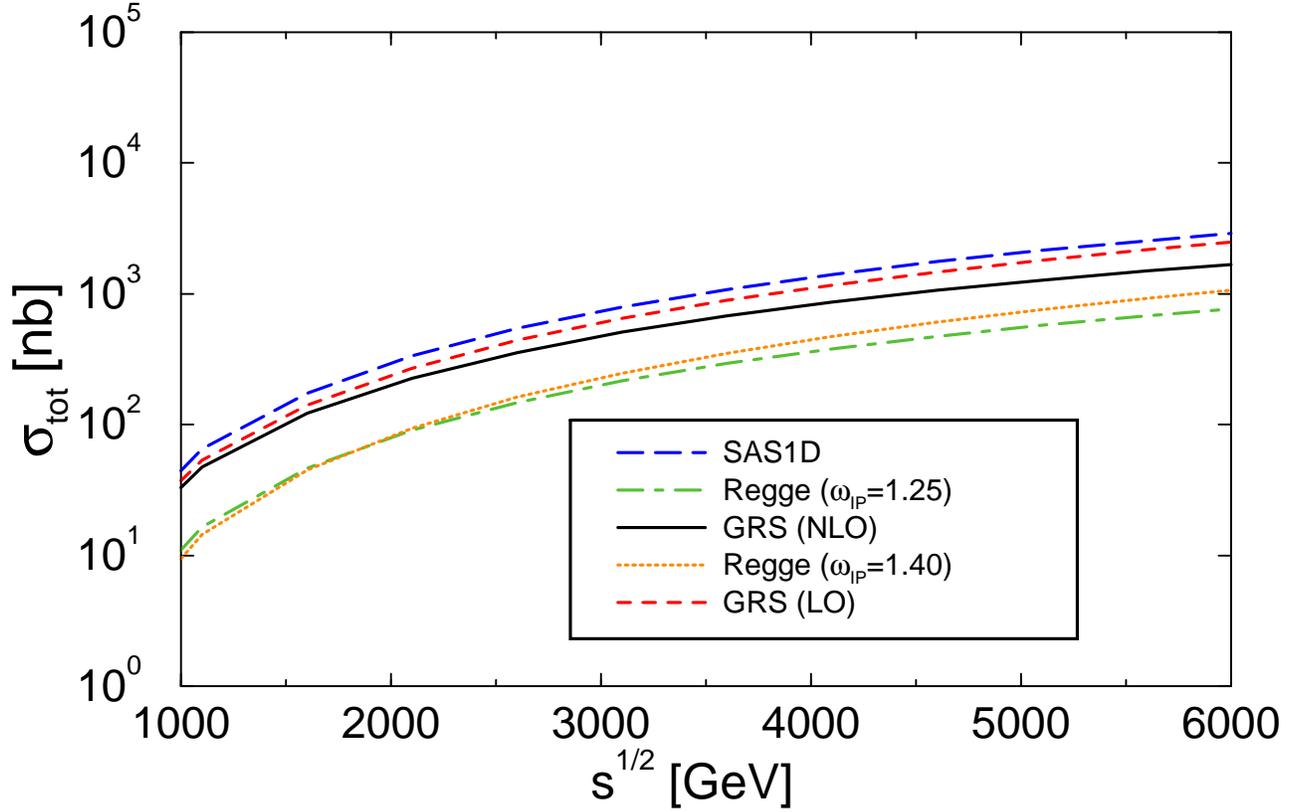,width=170mm}}
\caption{The total cross section for $\rho \,J/\Psi$ production in   two-photon ultraperipheral heavy ion collisions  as a function of the ion cms energy $\sqrt{s}$ (A = 208). The distinct curves correspond to the different parameterizations. }
\label{fig4}
\end{figure}

In Fig. \ref{fig4} are presented the results for the distinct  parameterizations discussed above, considering $A=Pb$. For sake of completeness, we also include the Regge ansatz with a higher Pomeron intercept, $\omega_{\pom}=1.4$, in lines of the studies on double $J/\Psi$ production \cite{vicmag}. A upper bound is given by the SaS1D parameterization and the lower one by the Regge ansatz with the lower Pomeron intercept $\omega_{\pom}=1.25$. At the LHC energies $\sqrt{s}=5500$ GeV, the cross section takes values between 680 nb and  2.6 mb, showing the process can be used to discriminate among parameterizations. It should be noted that the ultraperipheral cross section is dominated by not so high two-photons energies. Although the deviations among the models are quite sizeable at very high energies, as shown in Fig. \ref{fig2}, in the nuclear cross section such deviations are less sizeable. This is directly related to the effective two-photon luminosity, which peaks at smaller $W_{\gamma \gamma}$. Therefore, correct estimates should include careful treatment of the low energy threshold effects. The nuclear cross section is enhanced in relation to the $e^+e^-$ case, which provides values of hundreds of pb's in contrast with units of mb in the peripheral nuclear case.

\begin{figure}[t]
\centerline{\psfig{file=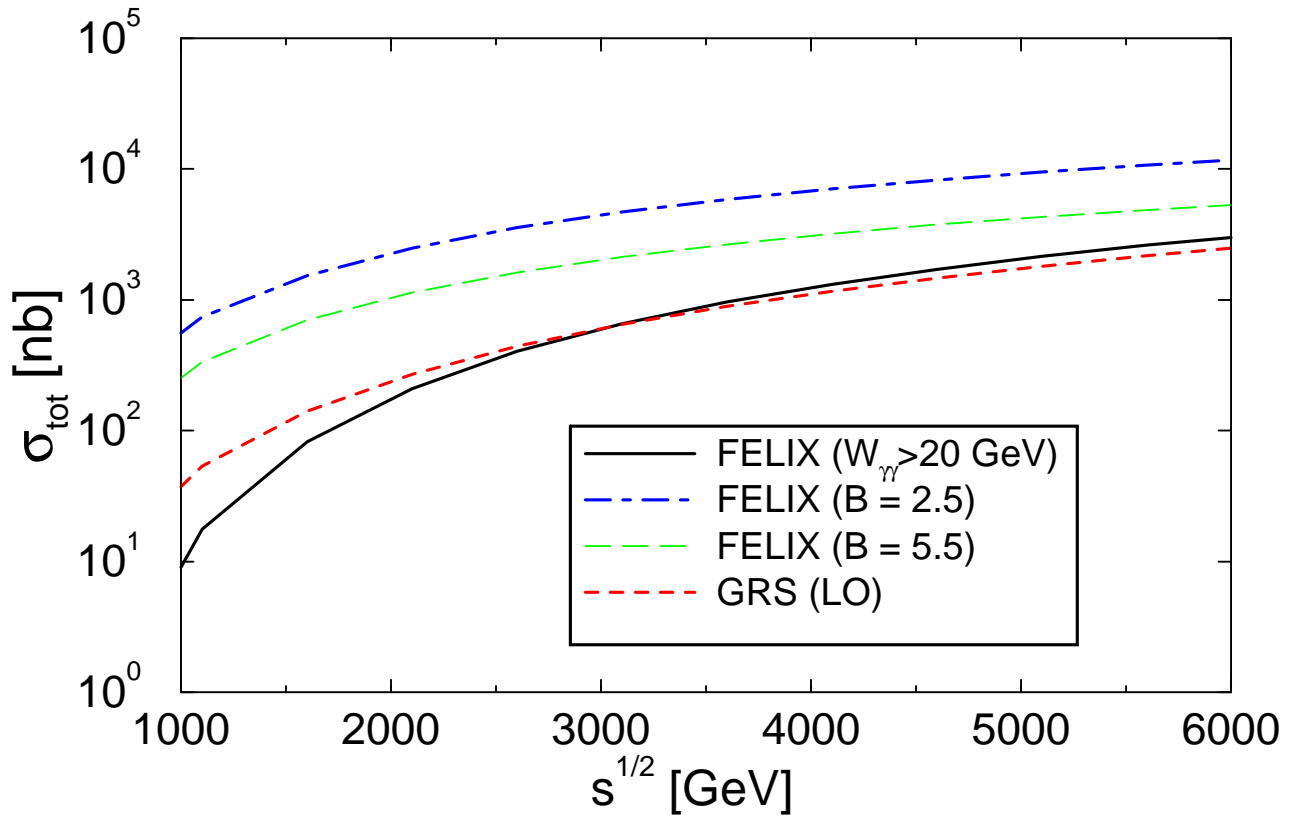,width=170mm}}
\caption{The same as Fig. \ref{fig4} considering the FELIX ansatz for the cross section with (solid curve) and without (dot-dashed curve)  the kinematical cut.  For comparison our prediction using  GRS(LO)  input is also presented.
 }
\label{fig5}
\end{figure}

In Fig. \ref{fig5} the results considering the FELIX ansatz  for the two-photon cross section and different values of the slope are presented. For comparison the predictions obtained with our approach and the GRS(LO) input are also shown. We see that the large difference between the GRS(LO)  and FELIX results for low energies in the two-photon case (See Fig. \ref{fig3})    implies that the values predicted by FELIX collaboration are ever larger than obtained using our approach, independently of the value of slope used. As a cross check, we also present in  Fig. \ref{fig5} the predictions for $\rho J/\Psi$ production in ultraperipheral heavy ion collisions obtained using the FELIX ansatz and the kinematical cut proposed in Ref.  \cite{felix}
. We have that our results agree with those presented in that reference.

As a by product we can use our results for double $J/\Psi$ production obtained in Ref. \cite{vicmag} and that presented above for  $\rho J/\Psi$ production to   estimate the double $\rho$ production in ultraperipheral heavy ion collisions, using the factorization theorem to relate the distinct cross sections. Given the assumption that single Pomeron exchange dominates, the following relation  among the distinct differential cross sections is predicted
\begin{eqnarray}
\frac{d\sigma}{dt} (\gamma \gamma \rightarrow V_1 V_1) = \frac{\left[\frac{d\sigma}{dt} (\gamma \gamma \rightarrow V_1 V_2)\right]^2}{ \frac{d\sigma}{dt} (\gamma \gamma \rightarrow V_2 V_2)} \,\,.
\label{facdsdt}
\end{eqnarray}
The usual approximation in the small $t$ region, $d\sigma /dt = A \, \exp\,(Bt)$, where $A = (d\sigma /dt)_{t=0}$, gives
\begin{eqnarray}
\sigma(\gamma \gamma \rightarrow \rho \rho) = \frac{(B_{\rho J/\Psi})^2}{B_{\rho \rho} \times B_{J/\Psi J/\Psi}} \times \frac{\left[\sigma (\gamma \gamma \rightarrow \rho J/\Psi)\right]^2}{\sigma(\gamma \gamma \rightarrow J/\Psi J/\Psi)}\,\,.
\label{facsig}
\end{eqnarray}
The slope of the diffractive peak for the $\gamma \gamma \rightarrow J/\Psi J/\Psi$ process was estimated in Ref. \cite{vicmag} as being $B_{J/\Psi J/\Psi} = 1/m_c^2$, where $m_c$ is the charm quark mass. Using the factorization theorem for the slopes, similarly to made in Eq. (\ref{facbs}), we estimate $B_{\rho \rho} \approx 12.1 \,$ GeV$^{-2}$.   
In the Tables \ref{tab1} and \ref{tab2} we present our predictions for the double $\rho$ production in $\gamma \gamma$ and ultraperipheral $AA$ collisions, considering different scenarios for the QCD dynamics. We can see that there is a large range of possible values for the cross section, which implies that future experimental data are essential to constraint the dynamics. It is important to salient that our predictions agree with those presented in Ref. \cite{felix}.

In Ref. \cite{klein} was emphasized that due to the large cross sections for vector meson production in photon-Pomeron interactions, the probability of having multiple interactions in a single nucleus-nucleus encounter is non-negligible. Therefore, it is possible that in double photonuclear interactions occur  the production of two mesons, generating a similar signal of the two-photon processes discussed here. There, the authors have presented some estimates of the cross sections for productions of vector mesons pairs in lead beams at LHC neglecting possible correlations between the multiple interactions (See Table V of Ref. \cite{klein}). Consequently, it is important to compare the predictions for the production of meson pairs in two-photon  and photon-Pomeron interactions. Our predictions indicate that the $\sigma (AA \rightarrow AA  \rho J/\Psi) \leq 10$ $\mu$b and  $\sigma (AA \rightarrow AA  \rho \rho) \leq 1250$ $\mu$b, independently of the scenario considered. These values are almost one order of magnitude below of the predictions presented in Ref. \cite{klein}, which implies that the rates of photonuclear multiple interactions would be an important background for the two-photon reactions, also for double meson production. In principle,  
an analysis  of the impact parameter dependence should allow to separate between the two
 classes of reactions, since two-photon interactions can occur at
 a significant distance from both nuclei, while a photonuclear
 interaction must occur inside or very  near a nucleus. We salient
 that the experimental separation between the two classes of
 processes is an important point, which deserves more studies.

As a summary, in this paper we have estimated the $\rho J/\Psi$ and $\rho \rho$ production in two-photon processes at ultraperipheral heavy ion collisions. We have verified that if these measurements are feasible, a discrimination among the current parameterizations for the gluonic content of the mesons is possible.  As a consequence, the analysis allows to constraint the partonic content  to  the photon structure function. Using the Pomeron-exchange factorization relations, the rates for double meson production were calculated and contrasted with the current predictions in the literature. The values found are in agreement with previous estimates. Concerning possible background processes, it was verified that the multiple meson production in photonuclear reactions constitutes the main source. The experimental separation between such processes should be taken into account, in particular focusing on impact parameter dependence and in rapidity distribution.

\section*{Acknowledgments}
This work was
partially financed by the Brazilian funding agencies CNPq and
FAPERGS. M.V.T.M. thanks the support of the  High Energy Physics Phenomenology 
Group at the Institute of Physics, GFPAE IF-UFRGS, Porto Alegre.

\newpage

\begin{table}[t]
\begin{center}
\begin{tabular}{||c|c||}
\hline
\hline
 SCENARIO  & $\sigma (\gamma \gamma \rightarrow \rho \rho)$ (pb)\\
\hline
BFKL(LO) + GRS(LO)  & $5 \times 10^3$ \\
\hline
BFKL(LO) + GRS(NLO) & 800  \\
\hline
BFKL(LO) + Regge ($\omega_{\pom} = 1.25$) & 50  \\
\hline
BFKL(MOD) + GRS(LO)  & $5 \times 10^5$ \\
\hline
BFKL(MOD) + GRS(NLO) & $80 \times 10^3$  \\
\hline
BFKL(MOD) + Regge ($\omega_{\pom} = 1.25$) & $5 \times 10^3$  \\
\hline
\hline
\end{tabular}
\end{center}
\caption{The double $\rho$ production cross sections in
$\gamma \gamma$ processes ($W = 100$ GeV). The notation BFKL (LO/MOD) corresponds to the treatment for the double $J/\Psi$ production in Ref. \cite{vicmag} and GRS (LO/NLO), Regge,  correspond to the treatment for the $\rho\,J/\Psi$ production presented here. } 
\label{tab1}
\end{table}

\newpage

\begin{table}[h]
\begin{center}
\begin{tabular}{||c|c||}
\hline
\hline
 SCENARIO  & $\sigma(A A \rightarrow A A \rho \rho)$ (nb)\\
\hline
BFKL(LO) + GRS(LO)  & $25 \times 10^3$ \\
\hline
BFKL(LO) + GRS(NLO) &  $4 \times 10^3$  \\
\hline
BFKL(LO) + Regge ($\omega_{\pom} = 1.25$) & 810  \\
\hline
BFKL(MOD) + GRS(LO)  & $125 \times 10^4$ \\
\hline
BFKL(MOD) + GRS(NLO) & $20 \times 10^4$  \\
\hline
BFKL(MOD) + Regge ($\omega_{\pom} = 1.25$) & $405 \times 10^2$  \\
\hline
\hline
\end{tabular}
\end{center}
\caption{The double $\rho$ production cross sections in
ultraperipheral heavy ion collisions at LHC ($\sqrt{s}=5500$ GeV)
for PbPb. Same notation as in the previous table.} \label{tab2}
\end{table}

\newpage

\newpage

\newpage

\newpage

\end{document}